\documentclass[12pt,preprint]{aastex}

\slugcomment{Draft as of \today}

\shorttitle{Gravitational Waves from 1987A}
\shortauthors{Santostasi, Johnson, Frank}

\newcommand{\etal}{{\it et al.~}}
\newcommand{\bfdelta}{\mbox{\boldmath{$\delta$}}}
\newcommand{\bfomega}{\mbox{\boldmath{$\omega$}}}

\def\spose#1{\hbox to 0pt{#1\hss}}

\def\lta{\mathrel{\spose{\lower 3pt\hbox{$\mathchar"218$}}
     \raise 2.0pt\hbox{$\mathchar"13C$}}}
\def\gta{\mathrel{\spose{\lower 3pt\hbox{$\mathchar"218$}}
     \raise 2.0pt\hbox{$\mathchar"13E$}}}

\received{2002 March 14}
\begin{document}

\title{ Detectability of Gravitational Waves from SN 1987A }

\author{Giovanni Santostasi, Warren Johnson \and Juhan Frank}
\affil{Department of Physics and Astronomy, Louisiana State University,
    Baton Rouge, LA 70803-4001}

\begin{abstract}
We discuss the potential for detection of gravitational waves from
a rapidly spinning neutron star produced by supernova 1987A
taking the parameters claimed by \citet{mid00} at face
value. Assuming that the dominant mechanism for spin down is
gravitational waves emitted by a freely precessing neutron star,
it is possible to constrain the wobble angle, the effective moment of 
inertia of the precessing crust and the crust cracking stress limit.
Our analysis suggests that, if the interpretation of the Middleditch 
data is correct, the compact remnant of SN 1987A may well 
provide a predictable source of gravitational waves
well within the capabilities of LIGO II. 
The computational task
required for the data analysis is within the capabilities of
current computers if performed offline and could be accomplished online
using techniques such as demodulation and decimation.

\end{abstract}


\section{Introduction}
\label{intro}

\citet{mid00} have claimed the likely first detection of
the compact remnant of supernova 1987A (hereafter SN 1987A). 
Through fast photometry of a small region around the supernova they
were able to find a modulated signal with a main frequency of 467.5 Hz and a
modulation period of about 1000 seconds. Assuming that this was the 
spin frequency of the presumed pulsar, and following the source between 
1992 and 1996, they were able to determine the
spindown of the pulsar and changes in the precession period. The observations
were complicated by times when the pulsation or the modulation were 
not visible
or not so evident. The pulsations seem to have disappeared completely since
1996. While astrophysically plausible explanations for the intermittency
of the signal can be devised appealing to the very complex nature of the
SN 1987A environment, the reality of a pulsar with the described
characteristics is at best very suggestive.

In this paper we will simply assume that the pulsar interpretation is
correct, adopt the parameters derived by Middleditch \etal 
at face value and derive some interesting implications for the detection
of gravitational waves from this source. We base our discussion
on simple free precessing neutron star models \citep{alp85, cut00, jan01}.

The general problem of emission of gravitational waves from rotating
and precessing neutron stars including pulsars and low-mass X-ray binaries
has been recently reviewed by \citet{jon01}. For the particular case of 
SN 1987A, while other authors \citep{cut00, jan01, nag01}
have also examined some of the consequences of 
the results of \citet{mid00}, none of these papers 
provides a precise calculation of the intensity and detectability of
gravitational waves from this source. The main aim of this paper is
to provide these estimates and to discuss the likelihood of detection
of gravitational waves from the hypoyhetical pulsar in SN 1987A
by LIGO I and II.  In Section \ref{strength} we estimate
 the time required to 
observe the signal with different types of
detectors using coherent integration techniques.
We show that, within a plausible range of
values of the moment of inertia $I_{0},$ 
the gravitational wave strain is big
enough to be detectable by LIGO II within integration times
ranging from days to months. Thus, if the interpretation of the 
periodicities in the optical observations is correct, 1987A 
{\em should} be a predictable source of gravity waves 
for ground based observatories. The computational requirements
for the data analysis discussed in Section \ref{templates}
are non-trivial but within the capabilities of modern computers.

\section{Summary of the Observations}
\label{observations}

\citet{mid00} discuss fast photometry observations 
of the remnant of the
supernova 1987A carried out at different times over the
period 1992--1996 from several observatories. During that time interval the
pulsar was detected several times at slightly different frequencies.
The power in the signal faded since 1993 and was last detected in February 
1996.
While they found ``no clear
evidence of any pulsar of constant intensity and stable timing,'' they did
find ``emission with a complex period modulation near the
frequency of 467.5 Hz - a 2.14 ms pulsar candidate''. They also point out
that: ``the frequency of the signals followed a consistent and
predictable spin-down ( $\sim $2-3 x 10$^{-10}$ Hz/s) over the several year
timespan. They find evidence for ``modulation of the 2.14 ms period with a
$\sim $1,000 s period which complicates its detection.''

The observed modulation of the 2.14 ms period can be interpreted as the effect 
of precession due to some deformation or crustal density
distribution which is not symmetric about the axis of rotation, including the case
in which the
precesing object itself possesses axial symmetry about a body axis which is not
aligned with the axis of rotation. In the absence of any external torques,
this situation is termed ``free precession". Classical mechanics tells us that
the ratio between the precession frequency and the rotation frequency is
proportional to the size of the deformation (e.g. \citet{mar95}). The size 
of the deformation and the frequency of rotation determine the rate of spin-down if 
the neutron star is assumed to lose energy mainly due to
gravitational radiation. 

A freely precessing neutron
star emits gravitational waves (e.g. \citet{zim79}, 
\citet{zim80}). 
Using the general relativistic energy loss
equation and the classical mechanics relationship between ellipticity, rotation
and precession frequency, we have that the spin down rate is proportional to the
square of the precession frequency under the assumption that all the energy is
lost due to gravitational back reaction. If an electromagnetic contribution to the
spin down rate is also present, this term would be independent of ellipticity
and would be approximately constant during the time span of the observations.
The data shown on Fig. 9 of \citet{mid00} are consistent with a linear
correlation between spin down rate and the square of the precession frequency 
going straight through the origin, i.e. with
zero contribution from magnetic dipole emission. Thus \citet{mid00}
conclude that the characteristics of the 
2.14 ms signature and its
$\sim $1,000 s modulation are consistent with precession and spindown via
gravitational radiation of a neutron star with effective non-axisymmetric
oblateness of $\sim $10$^{-6}$. We re-examine some aspects of this problem in 
Section \ref{model}.

\section{A Model for the Precessing Neutron Star}
\label{model}

\subsection{System Geometry}
\label{geometry}
Rotating neutron stars are often mentioned
as a possible continuous source of gravitational radiation. Usually what is
envisioned is that the star has an axissymetric deformation perpendicular to the
axis of rotation to allow for a changing mass quadrupole that will generate
gravity waves. Such a prolate or oblate star, tumbling about an axis 
perpendicular to the axis of symmetry, will emit gravity waves 
at twice the rotation frequency.

If the star is deformed on a axis that is
at any other angle with the rotation axis then it will precess as a spinning
top, and will emit at both twice the rotation frequency and at the rotation frequency. 
The
simplest situation is that the star is a rigid body and has just two non-equal
principal moments of inertia. We have then $I_{1}=I_{2}=I_{0}-\Delta I_{ \rm d}/3,$
$I_{3}=I_{0}+2/3\Delta I_{\rm d}$, so that $\Delta I_{\rm d} =  I_{3}-I_{1}$.
$I_{0}$ is the average value of \ the moment of inertia and $\Delta I_{\rm d}\ll I_0$.
A more complete and realistic model is
considered further below but the simplest case remains the basis for the discussion of
precessing neutron stars. The main equations are the same even in the more
realistic case with minor modifications. Figure 1 shows our convention in the
orientation of the important vectorial quantities involved in the
problem.

\clearpage

\newcommand{\mbf}{\mathbf}
\begin{figure} 
\plotone{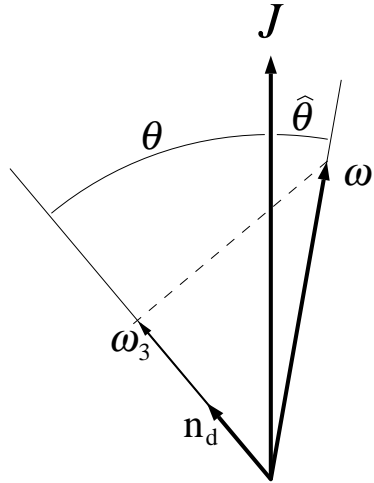}
\caption{The reference plane for a
freely precessing body. This diagram shows the respective orientation of the axis of
deformation ${\mbf n}_{\rm d}$, angular momentum ${\mbf J}$, and axis of rotation 
${\mbf \omega}$ of the star. The projection of the instantaneous angular velocity 
vector ${\mbf \omega}$ on to the symmetry axis ${\mbf n}_{\rm d}$ is indicated by 
${\mbf \omega}_3$. 
}
\label{fig1}
\end{figure}

\clearpage

We can
define the total moment of inertia as:
\begin{equation}
{\mbf I}=I_{0}\bfdelta + \Delta I_{\rm d}\left(
{\mbf n}_{\rm d}{\mbf n}_{\rm d}- \bfdelta/3\right) 
\label{itensor1}
\end{equation}
where
${\mbf n}_{\rm d}$ is a unit vector pointing along the body symmetry axis and
$\bfdelta$ is the unit tensor. Now define the ``ellipticity"
as a small quantity $\epsilon =\Delta
I_{\rm d}/I_{0}$, then classical mechanics implies
\begin{equation}
\epsilon =\frac{\Omega _{\rm p}}{\omega
_{3}}=\frac{\Omega _{\rm p}}{\omega \cos \gamma }\approx 
\frac{\Omega _{\rm p}}{\omega\cos \theta }
\label{eps}
\end{equation}
where $\theta $ is the angle
between the total angular momentum and the vector ${\mbf n}_{\rm d}$, $\gamma
=\theta +\widehat{\theta }\approx \theta .$ The angle between the rotation axis
and the angular momentum is a small quantity of order $\Delta I_{\rm d}/I_0.$ The
quantity $\Omega _{\rm p}$ is the precession frequency and $\omega $ is the rotation
frequency and $\omega _{3} $ its projection along the 3-axis that coincides in
this case with the axis ${\mbf n}_{\rm d}.$ We will proceed from the
assumption that we know the parameters \ $\Omega _{\rm p}$ and $\omega$
from the observations of SN 1987A by \citet{mid00}.

The observed modulation or precession period varied during the span of the
observations in the range from approximately 935 s to 1430 s, while $\omega$
or the spin period varied measurably but relatively little. Consequently the
observed variations in $\Omega _{\rm p}$ must be attributed to variations in 
$\epsilon$ or $\theta$ or both. Note, however, that the correlation between
$\dot\omega$ and $\Omega _{\rm p}$ claimed by Middleditch \etal requires
that $\theta$ remain constant. \citet{jan01} and \citet{jon01}
have claimed that it is not easy to imagine how significant variations in 
ellipticity can occur without affecting the wobble angle. We shall return to 
this question in Section \ref{constwobble} and argue that it is in fact 
unlikely that variations in epsilon can significantly change the wobble angle.

\subsection{Gravitational radiation caused
by misalignment}
\label{misalign}
To determine the size of deformation and
consequently the strain carried by the gravitational radiation on earth we need to
evaluate the wobble angle $\theta$.
This can be done assuming that the star
is losing energy solely through gravitational radiation. 
Then we can use the general
relativistic equation for the rate of energy emission by gravitational waves
\citep{zim79,zim80}:
\begin{equation}
\dot{E}=-\frac{2}{5}\frac{G}{c^{5}}\epsilon
^{2}I_{0}^{2}\omega ^{6}\sin ^{2}\theta \left( 16\sin ^{2}\theta +\cos
^{2}\theta \right), 
\label{dote}
\end{equation}
where the first and second terms in parenthesis represent the contributions 
at $2\omega$ and $\omega$ respectively.

If the only source of energy for this emission is the neutron
star's rotational energy reservoir $E=1/2$ $I\omega ^{2},$ we have then
$\dot{E}=\omega \dot{\omega I},$ so that
\begin{equation}
\dot{\omega
}=\frac{2}{5}\frac{G}{c^{5}}\epsilon ^{2}I_{0}\ \omega ^{5}\sin ^{2}\theta
\left( 16\sin ^{2}\theta +\cos ^{2}\theta \right).
\label{dotomega}
\end{equation}
Since the change in angular velocity
$\dot{\omega }$ is known from observations, it is possible to
solve equation (\ref{dotomega}) for the ellipticity as a function of the wobble 
angle for any given $I_0$. Figure
2 shows the relationship between the ellipticity $\epsilon $ and 
the wobble angle $\theta $ derived from equation (\ref{eps}) for
the observed range of
precession periods between 935 s and 1430 s (monotonically increasing curves).
 Figure 2 also shows the result of solving 
(\ref{dotomega}) 
(monotonically decreasing curves), 
for the range of observed values of $2\times 10^{-10} < \dot{\nu}< 3\times 10^{-10}$ Hz/s
for an arbitrarily chosen representative value of $I_0=10^{44}$ g cm$^2$ .

\clearpage

\begin{figure} 
\plotone{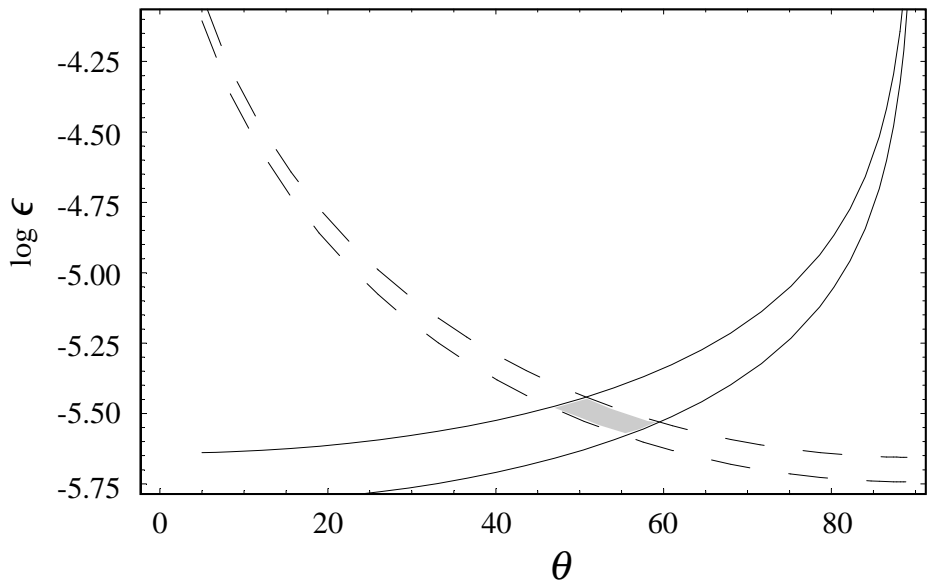}
\caption{
The general relativistic (dashed) and classical (solid)
relationships between $\epsilon$ and $\theta$ given the ranges of
observed values of spindown rate and precession frequency. The curves shown
correspond to 
$I_0=10^{44}$ g cm$^2$, a value intermediate between 
the minimum (just the crust precesses) and the
maximum (all the star is involved in the
precession). 
The possible solutions for the adopted value of the moment of inertia 
lie in the shaded region.
}
\label{fig2}
\end{figure}

\clearpage

The relativistic equation (\ref{dotomega}) depends on
the value $I_{0}$. This value is the average moment of the inertia of the part
of the star that actually participates in the precession. If the star has a crust
and a liquid interior then $I_{0}$ is the crust's moment of inertia and that of
any liquid coupled to the crust. In fact, part of the liquid should be stress
free and not influenced by the precession. So we can take $I_{0}$ to be an arbitrary
quantity equal or less than entire moment of inertia of the star
$I_{\rm star}=\frac{2}{5}MR^{2}=1.12\times 10^{45}\, {\rm g\ cm}^{2}M_{1.4}R_{6}^{2}$, 
where $M_{1.4}$ is the mass of the star in units of 1.4 solar masses and $R_{6}$ the
radius in units of 10$^{6}$ cm. If just the crust participates in the precession
then $I_{0}$ $\approx 1/100$ $I_{\rm star}$ according standard neutron star theory.
Now, the classical mechanics equation (\ref{eps}) and the
relativistic equation (\ref{dotomega}) have to be satisfied at the same time. 
This means
that for given observed $\Omega _{\rm p},\omega $, $\dot{\omega }$ and choice
of $I_{0}$ the functions have to meet at a point in the parameter space
$\epsilon -\theta .$ If we consider the moment of inertia the unknown parameter
of our problem we can determine which wobble angle the star should have
according the value of $I_{0}$. This is illustrated in Figure 3.
It seems that the 1987A remnant had some
relatively big and rapid changes in precession frequency during the first years of
observation. The astrophysical explanation for this could be a very active
dynamic environment in the young neutron star, that can bring abrupt changes in
the density of the crust, fractures and re-arrangement of surrounding material. We
already mentioned that Middleditch \etal find a power two relationship
between the observed change in $\dot{\omega }$ and $\Omega _{\rm p}.$ This
relationship holds exactly if we substitute 
equation (\ref{eps}) into (\ref{dotomega}), namely:
\begin{equation}
\dot\omega
=\frac{2}{5}\frac{G}{c^{5}}\frac{\Omega _{\rm p}^{2}}{\cos ^{2}\theta }I_{0}\
\omega ^{3}\sin ^{2}\theta \left( 16\sin ^{2}\theta +\cos ^{2}\theta \right),
\label{correlation}
\end{equation}
and require that $\theta$ remain constant while $\epsilon$ and hence $\Omega_{\rm p}$
vary.

\clearpage

\begin{figure} 
\plotone{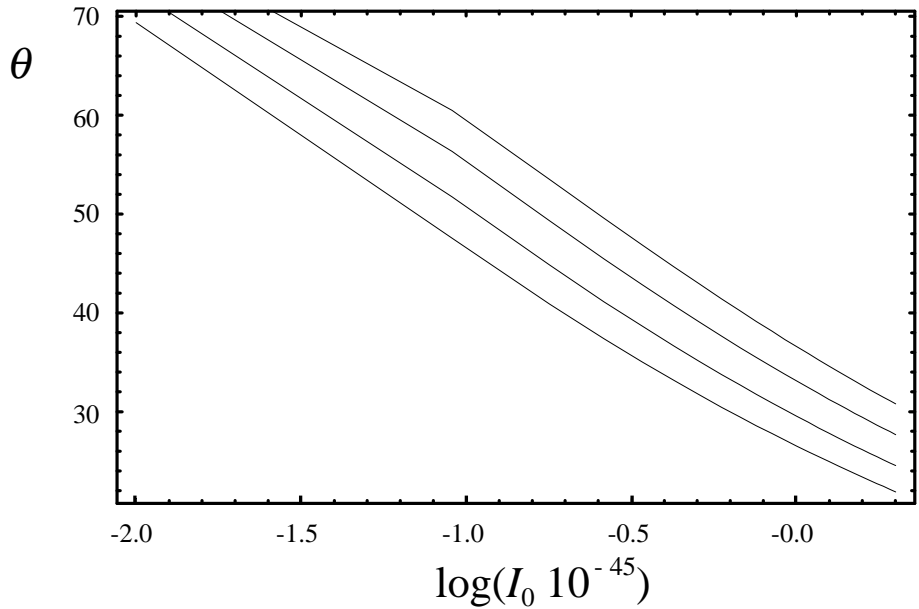}
\caption{
Wobble angle $\theta$ as a function of the moment of inertia
involved in the precession for the four possible combinations of
precession period (935 s and 1430 s) and spin down rate 
($2\times 10^{-10}$ Hz/s and $3\times 10^{-10}$ Hz/s).
The curves shown, from bottom to top, correspond to the following 
pairs of parameters: (935 s, $2\times 10^{-10}$ Hz/s),
(935 s, $3\times 10^{-10}$ Hz/s),  (1430 s, $2\times 10^{-10}$ Hz/s),
and (1430 s, $3\times 10^{-10}$ Hz/s), respectively.
}
\label{fig3}
\end{figure}

\clearpage

\subsection{The constancy of the wobble angle}
\label{constwobble}
\citet{mid00} display graphically the correlation between $\dot\omega$
and $\Omega_{\rm p}^2$. By reading off the values of these variables and applying 
the method and equations of Section \ref{misalign}, it is possible to determine the
values of the wobble angle $\theta$ required for each individual pair of values
$\dot\omega$ and $\Omega_{\rm p}$
for any assumed moment of inertia involved in the 
precession. This exercise reveals that despite variations of $\epsilon$ and $\Omega_{\rm p}$
exceeding a factor of 1.5, the wobble angle does not change by more than a couple of
degrees and appears consistent with remaining constant within experimental errors.

In the case of a freely precessing solid body, the wobble angle is largely determined 
by initial conditions: taking the principal axes introduced in Section \ref{geometry},
if the associated moments of inertia remain constant, then $\omega_3=\omega\cos{\gamma}$,
also remains constant (see Fig. \ref{fig1}). It is easy to generalize Euler's equations to 
the case in which the principal moments of inertia change due to unspecified
internal forces while the external torques vanish and the total angular momentum is conserved:
\renewcommand{\d}{{\rm d}}
\begin{eqnarray}
\frac{\d I_1 \omega_1}{\d t} &=& \omega_2\omega_3(I_2-I_3)\\
\frac{\d I_2 \omega_1}{\d t}  &=& \omega_3\omega_1(I_3-I_1)\\
\frac{\d I_3 \omega_1}{\d t}  &=& \omega_1\omega_2(I_1-I_2)\, .
\label{geneuler}
\end{eqnarray}
When the principal moments of inertia are all of the form $I_i=I_0 + \epsilon_i$, then 
clearly all the time derivatives are of order $\sim\epsilon_i$ and even if the given 
$\epsilon_i$ were to change by factors of a few, the result would be a small wobble of the 
tip of $\omega$ in the body frame. Therefore we conclude that while the variations 
probably detected by \citet{mid00} in
both $\epsilon$ and $\Omega_{\rm p}$ were significant, they do not imply any 
measurable change in $\theta$. Referring back to Fig. \ref{fig1} we see that 
$\widehat{\theta}$ may indeed change by amounts comparable to itself, but 
the wobble angle $\theta$ would change very little. This conclusion is contrary to 
what \citet{jan01} and \citet{jon01} have claimed regarding the 
wobble angle, and thus it makes more plausible that the remnant of SN 1987A is indeed
freely precessing while undergoing changes in  $\epsilon$ and $\Omega_{\rm p}$.

\subsection{A more realistic model: allowing for a
elastic crust and presence of a fluid interior}
\label{realistic}
The textbook discussion of a precessing
body assumes that the object is perfectly rigid. A more realistic neutron star will have a
more or less elastic shell, and a fluid interior. The fluid is supposed to be
composed of a electron-proton plasma and a neutron superfluid. The plasma fluid
interior can couple to the crust because of friction. Under these conditions the
system is not simply described by the rigid body model.

Usually the approach taken to explore the
properties of such more complicated systems is to understand the effect of one
additional complication at the time. The paper of \citet{cut00} addresses these
complications and shows how the more realistic model needs to be modified to
account for these complications. In this section we summarize these
results and apply them to the particular problem of the
detection of the 1987A remnant.

\subsubsection{The elastic crust}
\label{crust}

In the case of an elastic crust's shell we
have to write the moment of inertia as:
\begin{equation}
{\mbf I}=I_{0}\bfdelta + \Delta I_{\rm d}\left(
{\mbf n}_{\rm d}{\mbf n}_{\rm d}- \bfdelta/3\right) +
\Delta I_\omega\left(
{\mbf n}_{\omega}{\mbf n}_{\omega}- \bfdelta/3\right) 
\label{itensor2}
\end{equation}
this is the sum of a spherical part and two small
quadrupole contributions. 
The first term is the moment of the inertia of the undeformed
shell, in the absence of rotation. The second term is a deformation due to
Coulomb lattice forces and the third is the deformation due to centrifugal forces. The
vector ${\mbf n}_{\rm d}$ determines the axis of symmetry of the deformation
$\Delta I_{\rm d}$. The vector ${\mbf n}_{\omega }$ lies along the axis of
rotation and determines the direction the axis of symmetry of the centrifugal
deformation $\Delta I_{\omega }.$

The quantity $\Delta I_{\omega }$ is
caused by the deformation due to the centrifugal force, its value is determined
by: 
\begin{equation}
\frac{\Delta I_{\omega }}{I}=\frac{I_{0}^{2}\omega
^{2}}{4\left( A+B\right)} 
\end{equation}
where the constants $A$ and $B$ depend on
the particular stellar equation of state. The constant $A$ is on the order of
the gravitational binding energy and the constant $B$ is on the order of the
total electrostatic binding energy of the ionic crustal lattice. The quantity
$B$ is much smaller than $A $ so we can make the approximation:
\begin{eqnarray}
\frac{\Delta I_{\omega }}{I}\approx
\frac{I_{0}^{2}\omega ^{2}}{4A}\approx \frac{\omega ^{2}R^{3}}{GM}\nonumber \\
\approx 2.1\times 10^{-3}\left( \frac{f}{100\,{\rm Hz}}\right)
^{2}R_{6}^{3}/M_{1.4}
\end{eqnarray}
where $f$ is simply $\omega /2\pi$.

In the general situation of non parallel
${\mbf n}_{\rm d}$ and ${\mbf n}_{\omega}$, the body will
precess. As a consequence of ${\mbf n}_{\omega}$ being in the direction of
the rotation axis (at any given instant) the body will behave as a axysimmetric
top even if the body has a triaxial shape \citep{cut00}. 
The angular momentum of an arbitrary body ${\mbf J}={\mbf I} \bfomega$, 
with the inertia tensor ${\mbf I}$
given by equation (\ref{itensor2}), can be rewritten as 
${\mbf J}= {\mbf I}_{\rm eff} \bfomega$,
where ${\mbf I}_{\rm eff}=(I_{0}+2\Delta I_\omega/3)\bfdelta + 
\Delta I_{\rm d}\left({\mbf n}_{\rm d}{\mbf n}_{\rm d}- \bfdelta/3\right)$.
Thus in this case the three moments of inertia in the original body axes 
are: 
\begin{eqnarray}
I_{1} &=&I_{0}-\Delta I_{\rm d}/3+2\Delta I_{\omega}/3\nonumber\\
I_{2} &=&I_{1}\\
I_{3} &=&I_{0}+2\Delta I_{\rm d}/3+2\Delta I_{\omega}/3\nonumber
\end{eqnarray}
The main implication of this is that even in the
case of a elastic crust the star will still behave for what concerns precession
as a biaxial rigid object. The fundamental equation (2) holds for this situation
(with the appropriate inertia moments given above) and this means that the only
piece of the moment of \ inertia that contributes to precession is $\Delta
I_{\rm d}=I_{3}-I_{1}$.

\subsubsection{The presence of a fluid interior}
\label{fluid}
To further improve our model we consider the effects of the presence 
of a fluid interior. 
The shape of the cavity and the viscosity of the fluid contained are
important parameters. If the cavity is spherical, the presence
of the fluid in absence of viscosity has no influence on the 
precession. If the cavity is non-spherical, then there will be a reaction force
that is generated by the tendency of the fluid to assume axial symmetry
around the axis of rotation. The shell will be pushed by the fluid. This problem
is solved in the literature (Lamb 1932; Jones \& Andersson 2001)
under the assumptions of uniform fluid vorticity, 
small cavity ellipticity, and small wobble angle.  A small wobble angle is
adopted in the treatment given by \citet{lam32} only for mathematical convenience,
but this assumption can be safely relaxed as long as the ellipticity remains small 
without altering the result. The upshot is that the usual
precession equations described above are still valid. The only modification
to take into account is that $\Delta I_{\rm d}$ refers to the difference in moment of
inertia along the axis 1 and 3 of the whole star, and $I_{0}$ refers to the
average moment of the inertia of the shell only.

In the presence of friction between the
crust and a part of the interior fluid in contact with the crust we could have
some coupling between the motion of the crust and the core. It can be shown that
in the case of neutron stars the coupling is very weak and the core does not
participate in the precession. If there are frictional forces at work in the
interior of the star these will serve just to damp the free precession on time
scales between 400 and 10$^{4}$ precession periods \citep{alp85}.

\subsubsection{The problem of pinning}
\label{pinning}
 \citet{jan01} following and extending previous
work by \citet{sha77} conclude that the presence of pinning of the superfluid to the
crust, at least in the simplest possible configuration does not change the form
of the equations that describe the precession. The main modification required is
that the relevant {\it effective ellipticity} is generated by combination
of the lattice deformation and the moment of the inertia $I_{\rm SF}$ of the pinned
fluid, as in the following: 
\begin{equation}
\epsilon_{\rm eff}= \frac{\Delta I_{\rm d}}{I_{0}}+\frac{I_{\rm SF}}{I_{0}}
\end{equation}

The most common theories on pulsar
glitches give a precise prediction on the precession behavior in the presence of
pinning in a neutron star. The theories require at least a few percent of the
total moment of inertia of the star to be in the pinned superfluid. 
Current understanding of neutron star properties indicates that the moment
of inertia of the crust
is a few percent of the total moment of inertia of the
star. These considerations imply that: 
\begin{equation}
\epsilon_{\rm eff}=\frac{\Omega _{\rm p}}{\omega \cos \theta }=\frac{\Delta
I_{\rm d}}{I_{0}}+\frac{I_{SF}}{I_{0}}\approx 1,
\end{equation}
in the case of small
deformations $\Delta I_{\rm d}.$ The precession and rotation frequency should be
close in value if there is a sizable quantity of superfluid that is pinned to
the crust. These predictions are not confirmed by observations of the three
strong cases of precession in neutron stars: PSR\ B1642-03 \citep{jan01}, PSR\ B1828-11 
\citep{sta00} and the SN
1987A remnant, where the precession is on a time scale much longer than the
rotation. The conclusion is that if the free precession interpretation of the
modulation of the signal of these pulsars is correct, then there is 
almost no pinned superfluid in these stars (see \citet{link02} for further
discussion of this issue).

\section{The wobble angle and crust fracture}
\label{fracture}

Precession will cause the rotation axis of
the star to change its position relatively to the body frame. This means that
the centrifugal force distribution will be a function of position and time with 
a timescale on the order of the precession period. 
If the star has an elastic crust, then it will change its shape
in response to variations in the centrifugal force and
cause time dependent stresses in the crust. A simple order of magnitude 
estimate of the strain on the crust $\sigma$ due to precession yields:
\begin{equation}
\sigma \approx \left(\Delta I_{\omega}/I\right)\sin{\theta} \approx 0.046\sin{\theta}
\ R_{6}^{3}/M_{1.4}.
\end{equation}

Experiments with crystals suggest an
upper limit for the maximum possible strain sustainable by the crust before
breaking, i.e. $\sigma _{\max } \approx 10^{-2}.$ This implies that the possible
maximum wobble angle for our pulsar must satisfy: 
\begin{equation}
\theta\le\arcsin{(21.8\sigma _{\rm max})}, \nonumber
\end{equation}
which would require $\theta\le 13^\circ$ for $\sigma _{\max }=0.01$.
Since all of our estimated wobble angles are larger than 30$^\circ$, either 
$\sigma _{\max }$ is larger for the crust or the model is too simple
to account for the observations or our interpretation of the observations 
is incorrect.

\section{The strength of the radiation at earth}
\label{strength}

\citet{zim79} and \citet{zim80} treat the
case of a body with two distinct moments of inertia and obtain the following
expressions for the strain parameter $h$ of gravity waves from a neutron star at
a distance $r$ from Earth and average moment of inertia
$I_{0}$:
\begin{eqnarray}
\lefteqn{h_{+} =\frac{G}{c^{4}}
\frac{2I_{0}\omega^{2}\varepsilon\sin{\theta}}{r}\times }\nonumber \\
&\left[ (1+\cos^2{i})\sin{\theta} \cos{2\omega t}+ 
\cos i\sin i\cos \theta \cos{\omega t}\right]\nonumber\\
& \\
\lefteqn{h_{\times} =\frac{G}{c^{4}}
\frac{2I_{0}\omega^{2}\varepsilon\sin{\theta}}{r}\times }\nonumber \\
&\left[ 2\cos{i}\sin \ \theta \sin{(2\omega t)}+ 
\sin{i}\cos{\theta}\sin{\omega t}\right] \nonumber,
\label{hphx}
\end{eqnarray}
where $i$ is the unknown angle between the
angular momentum vector ${\mbf J}$ and the plane of the sky. 
It is important to notice that the time dependence of the
wave forms is sinusoidal with two main frequencies at $\omega $ and $2\omega.$
If the object was rotating along its symmetry axis it will emit just at a
frequency $2\omega $ (it will have also to be deformed along the axis
perpendicular to the rotation axis)$.$

We see that the frequency of rotation
$\omega $ is one of the important parameters that determine the strength of the
gravitational radiation on earth. We know from observation the value of the
rotation frequency to be $\omega =2\pi \ 467.5$ Hz. To determine what is the
strength of the radiation we need to know also the moment of inertia
$I_{0}$ involved in the precession and the wobble angle $\theta$. 
We showed previously that $\theta $ depends on how much of the moment of inertia of
the star is actually involved in the precession, as shown by the general
relativistic energy loss equation (\ref{dote}). When this relationship between $I_{0}$ and
$\theta$ is factored in the strain equations (17), we can determine the strength
of the gravitational radiation on earth as a function of the parameter $I_{0}.$
The result is shown in Figure 4.

\clearpage

\begin{figure} 
\plotone{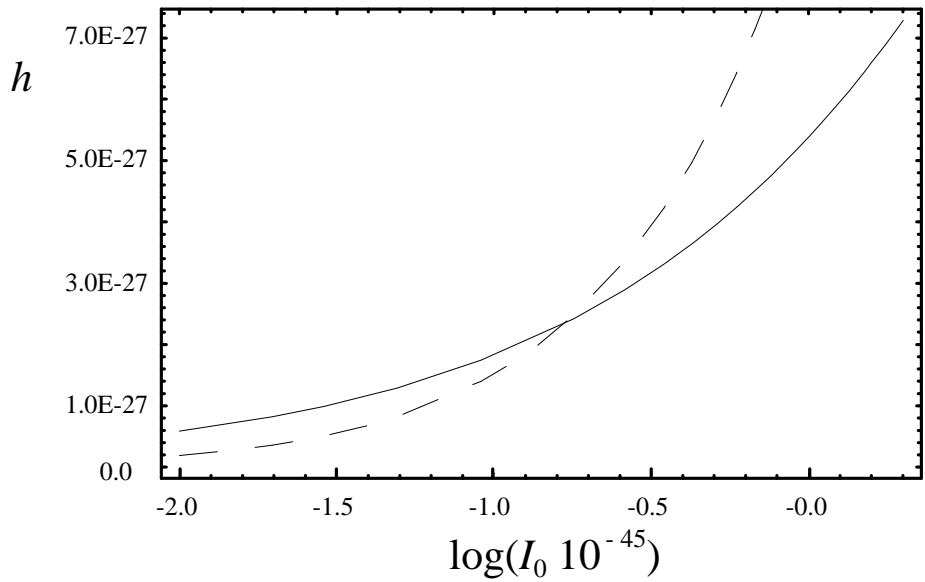}
\caption{
The amplitude of the strain at the Earth given by equation
(\ref{hphx}) -- leaving out dependences on $i$ and $t$ --
 as a function of the moment of inertia
involved in the precession, in units of $10^{45}$ g cm$^2$. 
The dashed curve corresponds to the
contribution from the $\omega$ term and the solid curve to the $2\omega$ term.
}
\label{fig4}
\end{figure}

\clearpage

These small values for $h$ may appear to
require an impossible level of sensitivity from the bar detectors or
interferometers existing today or soon available. It is important to notice that
the source is a continuous source of radiation, of which all the fundamental
parameters (besides the phase of the signal) are known. So it possible, in
principle, to integrate the detector data over a long period (even years) to
extract the signal from the incoherent noise. A detailed calculation of the
necessary integration time $\tau $ is required. To do so we use the following
equation:
\begin{equation}
h_{n}=\sqrt{S_h (f_0)}\sqrt{BW},
\end{equation}

This equation expresses the level of the
strain $h_{n}$ of the noise in the data from a detector with characteristic noise
spectrum $S_{h}$. The equation evaluates the value $S_{h}(f_{0})$ of the
spectrum at the precise frequency $f_{0}$ of the looked for gravitational wave
signal. The quantity $BW$ is the bandwidth of the periodic signal. From Fourier
Analysis theory in the case of a sinusoidal signal, the value of $BW$
=$\frac{1}{\tau },$ where $\tau $ is the observation or ``integration'' time. So
the required integration time is: 
\begin{equation}
\tau =
\left({\sqrt{S_{h}(f_{0})}\over h_{n}}\right)^{2}
\end{equation}

Now we require that the noise level in the
data from the detector be at least of the same size of the signal (it should
less, 4 times less for a 4 $\sigma $ confidence level in the statistics, for
example). A typical value for\ the noise strain $h_{\rm d}=\sqrt{S_{h}(f_{0})}$ in
the existing bar detectors such as the Louisiana State University's ALLEGRO or first
generation light interferometers \ as LIGO\ I, is currently of order $10^{-20}$.
The fully optimized LIGO I sensitivity is projected to attain a minimum noise strain
$h_{\rm d}\approx 10^{-22}$ in a couple of years (see Fig. 5).
We see from Figure \ref{fig4} that a typical value for the signal amplitude strain is
approximately $h_{s}=5\times 10^{-27}.$
So if $h_{n}$ is chosen to be $\approx
1/4\times h_{s}=1.25\times 10^{-27}$ then we get that:
\begin{eqnarray}
\tau = \left({\sqrt{S_{h}(f_{0})}\over h_{n}}\right)^{2}=
\left({10^{-20}\over 1.25\times 10^{-27}}\right)^{2} 
\left({h_{\rm d}\over 10^{-20}}\right)^2 {\rm s}\nonumber \\
\approx 2\times 10^{6} \left({h_{\rm d}\over 10^{-20}}\right)^2   {\rm yr}.
\end{eqnarray}
This is a time obviously too long to be useful, even with a fully optimized LIGO I.
Thus the remnant of supernova 1987A is
undetectable by the existing gravitational wave detectors, but may be 
detectable if the sensitivity of the detectors planned for the near future
reaches the estimated levels.

In fact preliminary estimates of the noise
spectrum of the second generation Laser Interferometer, LIGO II are very
promising (see Fig. 5). LIGO II will be built on the experience of the first LIGO and will be
a much better gravitational wave observatory. It will be on line in 4 or 5 years
from now. According to Fig. 5, there is a lowest point in the total strain-noise
(the sum of different kind of expected noises). This point is about
$h(f_{r})=$1.5$\times 10^{-24}$ /$\sqrt{Hz}$ at a frequency of 350 Hz. But the
LIGO\ II detector will be able to use narrow banding to shift this lowest point
in noise level to higher frequencies. For further discussion of noise levels 
expected in (Advanced) LIGO II, see \citet{abb02}.

\clearpage

\begin{figure}
\plotone{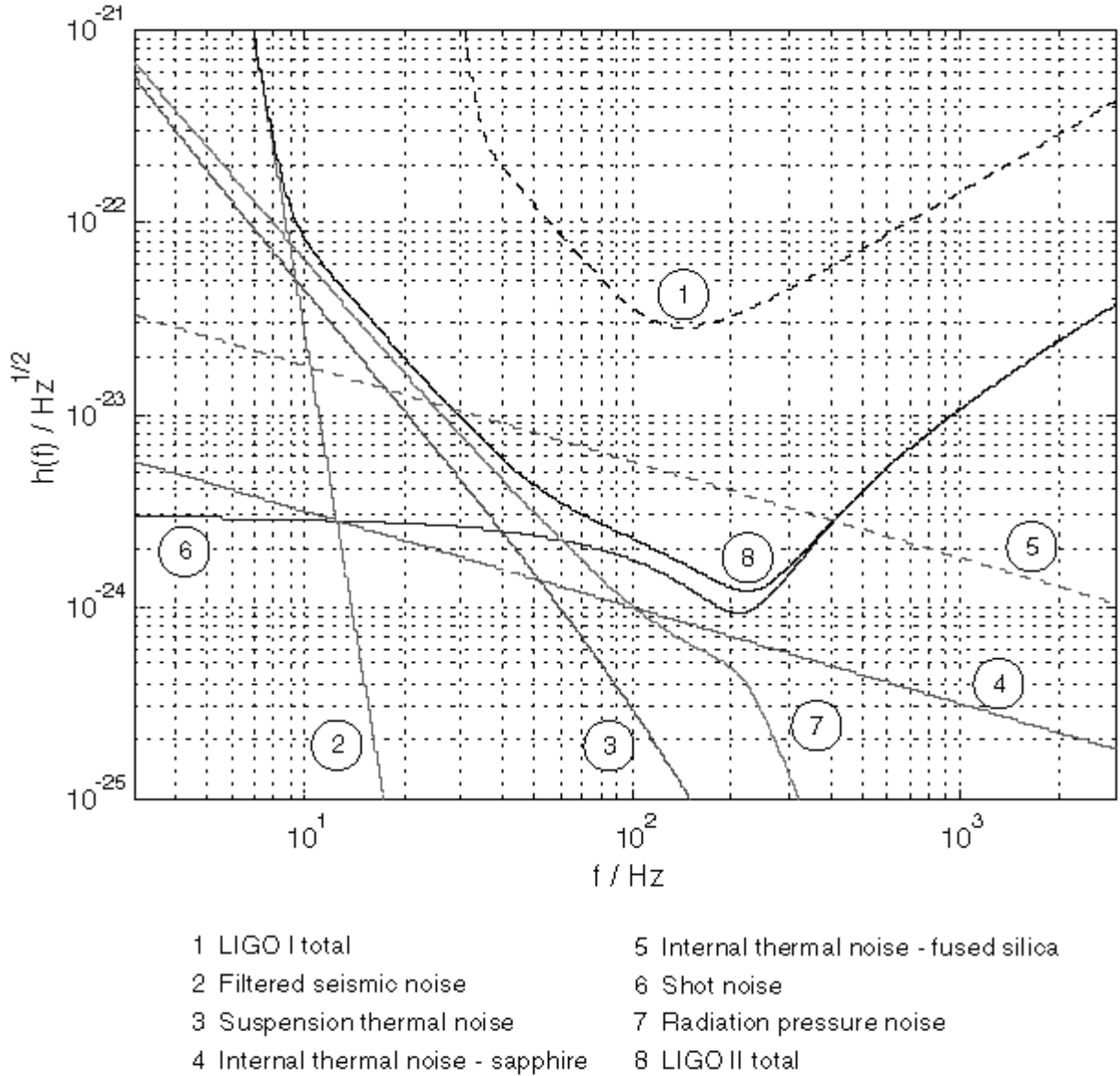}
\caption{
Preliminary estimates of the noise spectrum expected for LIGO II.
The contributions from various sources are shown separately and added
together (See http://www.ligo.caltech.edu/docs/M/M990288-A1.pdf).}
\label{fig5}
\end{figure}

\clearpage

So we could take this as the level of
noise at the frequency of emission of \ 1987A. Figure 6 shows the required time
of integration for LIGO\ II as a function of the parameter $I_{0}.$ We can see
that within a certain range of possible values of $I_{0}$ it will be possible to
detect the signal from the 1987A remnant in reasonable time. This time is
actually few days if most of the moment of the inertia participates in the
precession.

\clearpage

\begin{figure} 
\plotone{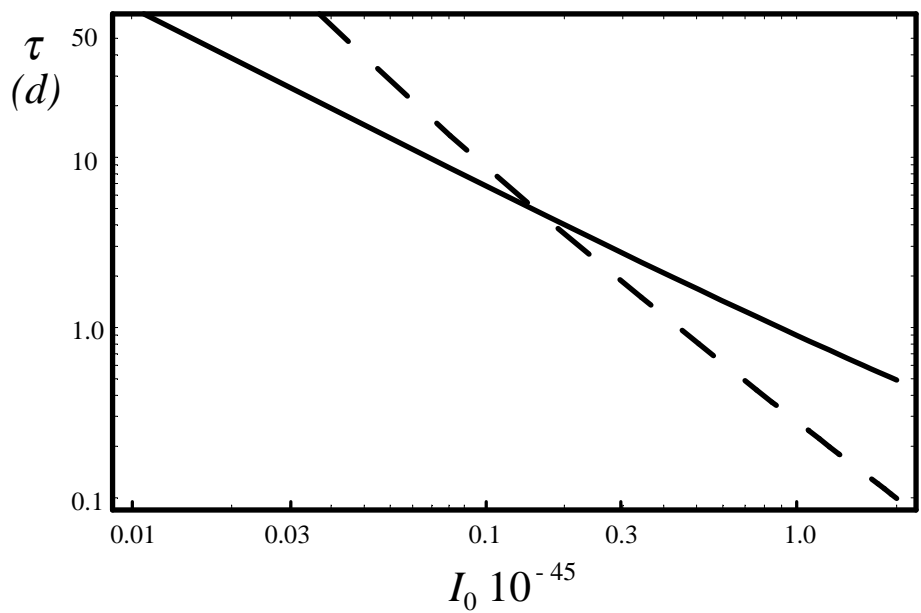}
\caption{
Estimated integration times for a 4 sigma detection by LIGO II for the signal at 
$\omega$ (dashed) and at $2\omega$ (solid) as a function of $I_0$ 
in units of $10^{45}$ g cm$^2$. The curves shown above have been estimated 
using the minimum strain level for an optimized LIGO II, $h_{\rm d} = 1.5\times 10^{-24}$.
For higher noise levels, the integration times scale as $(h_{\rm d}/1.5\times 10^{-24})^2$.
}
\label{fig6}
\end{figure}

\clearpage

\section{Data analysis and templates}
\label{templates}

The simple estimates for integration times in the previous section assume that
the signal is sinusoidal with constant frequency, which clearly is not the case
if our interpretation of the properties of the observed optical signal is correct. 
This raises the question 
of how detectable would be the signal if one allows for the observed
changes in the spin frequency, precession frequency and spin down rate. 
We show here that simple templates 
that describe the observed behavior can be constructed, and that the required 
number of such templates and the total computational effort needed to 
adequately keep track of the phase and to detect 
a signal with the presumed properties is 
within current computational capabilities
as long as there is phase
stability over a time series of length comparable or longer than the integration time.
Since the required integration times are on the order of 10--30 days, and the 
phase stability in the Middleditch data was comparable or better than that,
this requirement is likely to be satisfied. 

Following standard treatments we write the time-dependent frequency as a Taylor series 
\begin{equation}
\omega(t) =
\sum_{n=0}^\infty {\omega_n(0) t^n\over n!},
\label{templ1}
\end{equation}
where $\omega_n(0)$ indicates the $n$-th derivative at some arbitrary reference time
taken to be zero without loss of generality. A given choice of the parameters $\omega_n(0)$
constitutes a particular choice of template. Then the phase difference between two
different templates is 
\begin{equation}
\Delta\varphi (t) =
\sum_{n=0}^\infty {\Delta\omega_n(0) t^{n+1}\over (n+1)!}= 
\Delta\omega(0)\, t + {1\over2}\Delta\dot\omega(0)\, t^2 + \cdots\, ,
\label{templ2}
\end{equation}
where $\Delta\omega_n(0)$ represents the difference between the $n$-th derivatives
for a pair of templates. 

During the observations a typical value for the spin-down was 
$\dot f\sim 10^{-10}\, {\rm s}^{-2}$ and showing a secular decreasing trend.
Since the earliest opportunity for LIGO to observe this source is $T\sim$ 10 years
away, we take the uncertainty in the frequency to be on the order of 
$\dot f T \sim 3\times 10^{-2} \, {\rm s}^{-1}$ or  a bandwith of 
$BW=3\times 10^{-2}$ Hz. 
This is an estimate for the total range
of frequencies to be explored. The standard phase stability
requirement \citep{jakr00} $\Delta\omega\tau \lta \pi/4$ over the 
integration time, yields an estimate of how closely spaced the frequency 
templates have to be.  For $\tau\sim 10\,{\rm d}\sim 10^6\, {\rm s}$, this
argument yields $\Delta f = \Delta\omega/2\pi\sim 10^{-7}$ Hz. Consequently the 
total number of frequencies to be sampled is on the order of 
$BW/\Delta f\sim3\times 10^5 (\tau/10 {\rm d})$.

To calculate how many frequency derivative values need to be considered,
we estimate that the total range of values is comparable to $\dot f$ itself.
The phase stability requirement then yields the spacing between spin-down
values: $\Delta\dot f = \Delta\dot\omega/2\pi \lta \tau^{-2}/4 \sim 2.5 \times 10^{-13}$ Hz/s.
And therefore the total number of values of $\dot f$ to be sampled is 
approximately $\dot f/\Delta\dot f \sim 400 (\tau/10 {\rm d})^2$.

Finally, the total number of two-parameter templates $N_f$ we require is
 given by the simple product of the number of frequency values 
times the number of frequency derivative values: 
$N_f \sim 1.2\times 10^8(\tau/10 {\rm d})^3$.
The total number of floating-point operations required to carry out the search of 
these templates over an integration time $\tau$ is approximately given by the 
formula \citep{br98}
\begin{equation}
N_{\rm fpo} = 6 f N_f \tau [\log_2{(2 f \tau) +1/2]}\, ,
\label{flops}
\end{equation}
where $f$ is the maximum frequency to be searched (say 500 Hz). 
With the values derived above,
this yields a total computational load of 
approximately $1.1 \times 10^{19}(\tau/10 {\rm d})^4$ 
floating-point 
operations, which would require 3 months of calculations for a Teraflop machine.
While this load is not trivial, it can be achieved by either processing the data
offline or using a machine clocking at least 
$11 (\tau/10 {\rm d})^3$ Teraflops for online processing. 
However, the above estimate is an upper limit that makes little
use of our prior knowledge of the expected frequency and frequency range 
of the signal. We need only to search over the $BW$ of $3\times 10^{-2}$ Hz, whereas the
standard argument above assumes we are searching for signals over the entire
band from 0 to 500 Hz. The
computational task can be significantly reduced by first `demodulating' or filtering 
the signal to the bandwith $BW$ estimated above and then `decimating' or reducing the
signal sampling rate to the bandwith. This technique cuts the processing rate 
essentially by a factor $BW/f\sim 6\times 10^{-5}$ to approximately 
$0.7 (\tau/10 {\rm d})^3$ Gigaflops,
well within the capabilities of current computers.

\section{Conclusions}
\label{conclude}

In this paper we discussed the
implications of the observation of a precessing neutron star in the remnant of
supernova 1987A for gravitational wave detection. We used the observed data on
rotational velocity, spin down and precession rate to determine the value of the
possible asymmetric deformation that causes the precession.

To estimate the size of
deformation it is important also to determine the wobble angle 
between the axis perpendicular to the deformation and the rotation axis.
General relativity gives us an equation of the loss of energy, trough gravity
waves. Knowing the rate of spin down, the rotation frequency and the precession
frequency allows us to find the wobble angle. This is possible under the
assumption that the main mechanism for the loss of rotational energy is due to
emission of gravitational radiation. Once we know the wobble angle, 
we can calculate the
strength of the radiation on earth. In fact the value of the dimensionless
strain parameter $h$ depends on the value of the wobble angle quite strongly.

Our discussion shows that even with a more a realistic model
of a precessing neutron star that takes in consideration the presence of a
crust, with a certain elasticity and the eventual presence of a fluid interior
the precessional behavior is similar to that of the simple biaxial model.
The ratio of precesion frequency and spin frequency determines the order of 
magnitude of the ellipticity, but a complete solution requires an estimate of 
the wobble angle. The preceding discussion shows that it is possible to obtain 
such self-conssitent models as a function of essentially one parameter:
the moment of inertia $I_{0}$ that is involved in the precession. Given it, 
the observational data allow to determine the wobble angle, the size of
deformation and consequently the strength of the radiation on earth.

We saw in the previous section
that to avoid crust breaking the wobble angle has to be relatively small. 
In fact, formally, even the smallest wobble angle among the possible range 
of solutions violates the maximum crustal strain. Given the uncertainties in the 
model and in the interpretation of the data, we conclude that even if the
limits on the maximum strain $\sigma_{\rm max}$ are relaxed, any viable solution
is likely to have a wobble angle near the small end of the range and consequently
the moment of inertia must be near the high end of its range. 
In turn this means that a short integration time on the order of days
is required to
observe with confidence the gravity wave signal from SN1987A using advanced
detectors as LIGO\ II. Unfortunately the presently generation of 
detectors such as the resonant bars and LIGO I would require 
observation times of the order
of a million years 
to extract the signal from the noise.
Thus if the precession interpretation is correct,
the SN 1987A remnant would be among the best candidates for
{\em a search for a}
continuous source of gravitational waves. In any case, it is clear that 
a targeted search for gravitational waves from this source is worthwhile
since both detection and absence of detection over a relatively short time 
will yield interesting constraints on models for precessing neutron stars.

\acknowledgements
This research was supported in part by NSF's Experimental Gravity Program
grant 9970742 and by grants AST9720771 and NAG5 8497 to LSU.

\end{document}